\documentstyle[12pt]{article}
\font\titlefont=cmbx10 scaled \magstep3
\begin{document}
\input{epsf}

\begin{flushright}
\vspace*{-2cm}
quant-ph/9804056  \\
CBPF-NF/007/98       \\  
April 22, 1998   \\
Revised June 2, 1998
\vspace*{2cm}
\end{flushright}

\begin{center}
{\titlefont Vacuum Energy Density near \\
\vskip 0.15in
Fluctuating Boundaries }
 
\vskip .3in
L.H. Ford\footnote{email: ford@cosmos2.phy.tufts.edu} \\
\vskip .1in
Institute of Cosmology,
Department of Physics and Astronomy\\
Tufts University\\
Medford, Massachusetts 02155\\
\vskip .2in
N.F. Svaiter\footnote{email:nfuxsvai@lafex.cbpf.br} \\ 
\vskip .1in
Centro Brasileiro de Pesquisas Fisicas-CBPF \\ 
Rua Dr. Xavier Sigaud 150\\ 
Rio de Janeiro, RJ 22290-180, Brazil \\
\end{center}

\vskip .2in
\begin{abstract}
The imposition of boundary conditions upon a quantized field can lead to
singular energy densities on the boundary. We treat the boundaries as quantum
mechanical objects with a nonzero position uncertainty, and show that the 
singular energy density is removed. This treatment also resolves a long standing
paradox concerning the total energy of the minimally coupled and conformally
coupled scalar fields.
\end{abstract}
\vskip .1in
 PACS categories: 03.70.+k, 12.20.Ds, 04.62.+v.
\newpage

\baselineskip=14pt

\section{Introduction}
\label{sec:intro}

 It is well-known that boundary conditions imposed upon quantum fields may lead
to divergent expectation values for local observables. A simple example is
a massless, minimally coupled scalar field $\varphi(t,{\bf x})$ which vanishes 
on the $z = 0$ plane;
\begin{equation}
\varphi \Bigl|_{z=0} = 0 \, .  \label{eq:phi=0}
\end{equation}
 One finds \cite{fulling} that the renormalized expectation values of 
both $\varphi^2$
and of the energy density $T_{tt}$ diverge as $z \rightarrow 0$. Specifically,
\begin{equation}
\langle \varphi^2 \rangle = - \frac{1}{16\, \pi^2\, z^2}  \label{eq:phisq}
\end{equation}
and 
\begin{equation}
\langle T_{tt} \rangle = - \frac{1}{16\, \pi^2\, z^4} \,.  \label{eq:Ttt}
\end{equation}
(Units in which $\hbar = c = 1$ will be used throughout this paper. The
metric tensor is taken to be $\eta_{\mu\nu} = diag(1,-1,-1,-1)$.)
The stress energy tensor of the massless minimally coupled scalar field 
is given by
\begin{equation}
T_{\mu\nu} = \partial_\mu \varphi \, \partial_\nu \varphi - \frac{1}{2} \,
\eta_{\mu\nu}\, \partial^\alpha \varphi \, \partial_\alpha \varphi \, .
                                                   \label{eq:Tmunu}
\end{equation}
Similar divergences occur in the expectation values of the squared electric 
or magnetic fields, $\langle {\bf E}^2 \rangle$ or $\langle {\bf B}^2 \rangle$ 
near a perfectly conducting plane, although in this case the local energy
density remains finite. When the conducting boundary is curved, then the 
energy density  diverges on the boundary \cite{DC}. 

Furthermore, there is  a puzzling discrepancy
between the Casimir energy for a minimal scalar field computed as the
renormalized expectation value of the Hamiltonian and as a spatial integral
of $\langle T_{tt} \rangle$. Consider the case of two parallel plates with
separation $L$ on which the field vanishes. If we first form the  Hamiltonian
operator $H = \int T_{tt}\, d^3x$, the result is the same for both the
minimal and conformally coupled fields. This follows from the fact that the 
stress tensor for the conformal field,
\begin{equation}
\Theta_{\mu\nu} = \partial_\mu \varphi \, \partial_\nu \varphi - \frac{1}{2} \,
\eta_{\mu\nu}\, \partial^\alpha \, \varphi  \partial_\alpha \varphi 
-\frac{1}{6} \Bigl[ \partial_\mu (\varphi \, \partial_\nu \varphi)
+ \partial_\nu (\varphi \, \partial_\mu \varphi) 
-2 \eta_{\mu\nu}\, \partial^\alpha (\varphi \, \partial_\alpha \varphi) \Bigr]
         \, ,                                          \label{eq:Tmunucon}
\end{equation}
differs from that for the minimal field, Eq.~(\ref{eq:Tmunu}), by a total
derivative term which integrates to zero. If we  find the  renormalized 
expectation value of $H$, the energy per unit area is found to be 
$- \pi^2/(1440\, L^3)$. However, if we attempt to compute this energy per 
unit area as $\int_0^L \langle T_{tt} \rangle \, dz$, the result is 
divergent. This discrepancy has led some authors \cite{KCD}  to postulate
the existence of a singular surface energy density, which would render the latter 
expression finite and equal to the former result. (Note that the surface energy 
densities which are of concern here are distinct from the surface-area-dependent
terms in the regularized Casimir energy which can arise in particular 
regularization methods \cite{SS1,SS2}.)

In curved or topologically nontrivial spacetimes, it is also possible for the
renormalized expectation value of the stress tensor to diverge on particular
boundaries. An example is the Boulware vacuum state in Schwarzschild spacetime,
for which the stress tensor diverges on the event horizon \cite{Unruh,visser}. 
This divergence
is usually interpreted as indicating that this is not a physically realizable
state. Other examples of divergent stress tensors include Misner space, where
the divergence occurs on the Cauchy horizon \cite{HK}. In this and similar examples,
one is tempted to resolve the problem by regarding the spacetime itself to
be unphysical. Indeed, this philosophy is the basis of Hawking's Chronology
Protection Conjecture \cite{HCP}, 
which argues that closed timelike curves are 
prohibited by the effects of divergent energy densities which would otherwise
appear on the chronology horizon (the boundary between a region containing
closed timelike curves and one without such curves). 

  One may understand why the imposition of a boundary condition such as
Eq.~(\ref{eq:phi=0}) on a quantum field can result in infinities. In the case
of $\langle \varphi^2 \rangle$, renormalization means taking the difference 
of the  expectation value in the presence of the boundary and in its absence.
Normally, this removes the infinite part and leaves a finite remainder.
However, on the boundary the formal expectation value of 
$\langle \varphi^2 \rangle$ is finite, so the subtraction results in an infinite
difference.  We can also understand why quantities such as 
$\langle \dot\varphi^2 \rangle$ and $\langle T_{tt} \rangle$ become
infinite on the boundary. The field $\varphi$ and its time derivative 
$\dot\varphi$ are conjugate variables which satisfy an uncertainty relation.
If $\varphi$ is precisely specified, $\dot\varphi$ is completely
indeterminate, and  $\langle \dot\varphi^2 \rangle$ and thus 
$\langle T_{tt} \rangle = \frac{1}{2}\langle \dot\varphi^2 + 
({\bf \nabla}\varphi)^2 \rangle$ are infinite. 
A state in which $\varphi$ is precisely
determined at a point has infinite energy density at that point for essentially
the same reason that a position eigenstate in single particle quantum mechanics
has infinite energy. 
 
    In the case of material boundaries, such  metal plates, infinite values
of $\langle {\bf E}^2 \rangle$ or other observables are presumably avoided
because such boundaries are not perfect reflectors at all frequencies. A
metal plate is a good reflector of electromagnetic waves at frequencies below
the plasma frequency, but becomes relatively transparent at higher frequencies.
Such a high frequency cutoff seems not to be available when the ``boundary''
is a feature of the spacetime structure. 

    The purpose of the present paper is to explore an alternative mechanism
for introducing a cutoff which removes singular behavior on boundaries. This
is to allow the position of the boundary to undergo quantum fluctuations. 
One might expect that such fluctuations will smear out the contributions of
the high frequency modes without the need to introduce an explicit  high 
frequency cutoff.

\section{$\langle \varphi^2 \rangle$ near a Single Plate}
\label{sec:phisq}
 
   Let us consider a plane boundary located at $z=q$. If we impose the
boundary condition on a massless quantized scalar field 
 $\varphi$ that it vanish on this boundary, the appropriate
two-point function may be constructed as an image sum. The result is
\begin{equation}
\langle \varphi(x) \varphi(x') \rangle = G(x,x') =  G_0(x,x') +  G_R(x,x') \, ,
\end{equation}
where 
\begin{equation}
G_0(x,x') = - \frac{1}{4\pi^2 (\Delta t^2 - \Delta {\bf x}^2) } \label{eq:gf}
\end{equation}
is the empty space two-point function, with $\Delta t = t-t'$, 
$\Delta {\bf x}^2 = |{\bf x} - {\bf x'}|^2$,
and
\begin{equation}
G_R(x,x') = \frac{1}{4\pi^2 \bigl[\Delta t^2
  - \Delta {x}^2 - \Delta {y}^2 - (z+z'-2q)^2 \bigr] } \, .  \label{eq:rgf}
\end{equation}
The full two-point function, $ G(x,x')$, vanishes whenever $z=q$ or $z'=q$.
The renormalized expectation value of $\varphi^2$ is given by the coincidence
limit of the renormalized two-point function, $ G_R(x,x')$,
\begin{equation}
\langle \varphi^2 \rangle = G_R(x,x) = - \frac{1}{16\, \pi^2\, (z-q)^2}
                                                          \label{eq:phisq2}
\end{equation}
and is singular at $z=q$. 

   We now wish to allow the position variable $q$ to fluctuate. This will occur
if the mirror is treated as a quantum object with a wavefunction $\psi(q)$, and
hence a position probability distribution of 
\begin{equation}
f(q) = |\psi(q)|^2 \, ,
\end{equation}
where
\begin{equation}
\int^\infty_{-\infty} f(q)\, dq = 1 \,.
\end{equation}
The average over position of a function $H(q)$ becomes
\begin{equation}
\langle H \rangle = \int^\infty_{-\infty} H(q)\, f(q)\, dq \,.
\end{equation}
Thus to find the mean value of $\varphi^2$, we need to calculate 
$\langle  G_R \rangle$. This is most easily done by expressing $ G_R$ in
a Fourier representation, and then averaging the $q$-dependence of the mode 
functions:
\begin{equation}
\langle  G_R(x,x') \rangle = - \frac{1}{2(2\pi)^3} \, {\rm Re} \int 
\frac{d^3 k}{\omega} \, e^{i{\bf k}_t \cdot ({\bf x}_t - {\bf x}'_t)}
\, e^{-i \omega (t-t')} \, e^{i k_z (z+z')} \, 
\langle e^{-2i k_z q} \rangle 
\, ,                   \label{eq:G_Rav}
\end{equation}
where ${\bf k}_t$ and ${\bf x}_t$ denote the components of ${\bf k}$
and ${\bf x}$, respectively, in directions parallel to the plate.

  To proceed further, we must specify the probability distribution, $f(q)$.
A convenient choice is a Gaussian peaked about $q=0$, 
\begin{equation}
f(q) = \sqrt{\frac{\alpha}{\pi}} \, e^{-\alpha q^2} \,,
\end{equation}
which leads to
\begin{equation}
\langle e^{-2i k_z q} \rangle = e^{-2 k_z^2 \langle q^2 \rangle} \, ,
                                                  \label{eq:exp}
\end{equation}
with 
\begin{equation}
\langle q^2 \rangle = \frac{1}{2 \alpha} \, .
\end{equation}
This probability distribution is the appropriate one to describe, for example,
a plate in the ground state of a harmonic potential. Note that 
Eq.~(\ref{eq:exp}) is equivalent to the result which one obtains when
taking the vacuum expectation value of the complex exponential of a free quantum
field. (See, for example, Eq.~(8) of Ref.~\cite{F95}.) This is to be expected,
as a free quantum field in the vacuum state is equivalent to an infinite
collection of harmonic oscillators in their ground states. 

    If we use Eq.~(\ref{eq:exp}) in Eq.~(\ref{eq:G_Rav}), set 
${\bf x}' ={\bf x}$, and use 
\begin{equation}
\int d^3 k = 2\pi \int^\infty_{-\infty} d k_z
\int^\infty_{|k_z|} d \omega \omega \,,
\end{equation}
for integrands independent of the azimuthal angle, the result is
\begin{equation}
\langle  G_R(t,t') \rangle = \frac{1}{2(2\pi)^2 (t-t')}\, {\rm Re} \left[ i
\int^\infty_{-\infty} d k_z \, e^{i (2 z  k_z+ (t-t')| k_z| )} \,
 e^{-2 k_z^2 \Delta^2}
 \right]   \, ,
\end{equation}
where 
\begin{equation}
\Delta = \sqrt{\langle q^2 \rangle}
\end{equation}
is the root-mean-squared displacement of the mirror. This integral may be 
performed in terms of the error function $\Phi$ to be
\begin{eqnarray}
\langle  G_R(t,t') \rangle &=& - \frac{i \sqrt{2\pi}}{8 (2\pi)^2 (t-t') \Delta}
  \left[ e^{-(2z -t+t')^2/(8\Delta^2)} \, 
\Phi\left(i \,\frac{t-t'-2z}{2\sqrt{2}\, \Delta}\right) \right. \nonumber \\
&+& \left. e^{-(2z +t-t')^2/(8\Delta^2)} \, 
\Phi\left(i \,\frac{t-t'+2z}{2\sqrt{2}\, \Delta}\right) \right] \,.
\end{eqnarray}
(Here and at many other places in this paper, the calculations  
were performed with the aid of the symbolic algebra program MACSYMA.)
This quantity is finite in the limit that $t' \rightarrow t$ :
\begin{equation}
\langle  G_R \rangle = \frac{\sqrt{2} \, z}{32 \sqrt{\pi^3} \Delta^3} \;
   e^{-z^2/(2\Delta^2)} \,i\,  \Phi\left(i \frac{z}{\sqrt{2} \Delta}\right)
\; + \; \frac{1}{16 \pi^2 \Delta^2} \,.
\end{equation}
Note that this expression is real, as may be seen from the fact that
\begin{equation}
\Phi(ix) = -\Phi(-ix) = \frac{2i}{\sqrt{\pi}} \, \int^x_0 e^{u^2} du \,.
\end{equation}

The quantity $\langle  G_R \rangle$ is $\langle \varphi^2 \rangle$ in the 
presence of position
fluctuations, and is finite for all $z$. For large $z$, we have
\begin{equation}
\langle  G_R \rangle \sim - \frac{1}{16\, \pi^2\, z^2} 
- \frac{3 \Delta^2}{16\, \pi^2\, z^4} +\cdots \,, \quad  
z \gg \Delta \, ,
\end{equation}
thus recovering the usual form far from the mirror. As $z \rightarrow 0$,
\begin{equation}
\langle  G_R \rangle \rightarrow \frac{1}{16\, \pi^2\, \Delta^2}\, ,
\end{equation}
and is hence finite. For intermediate values of $z$, 
$\langle  G_R \rangle$ may be computed numerically, and is 
depicted in Fig. 1.

\begin{figure}
\begin{center}
\leavevmode\epsfysize=8cm\epsffile{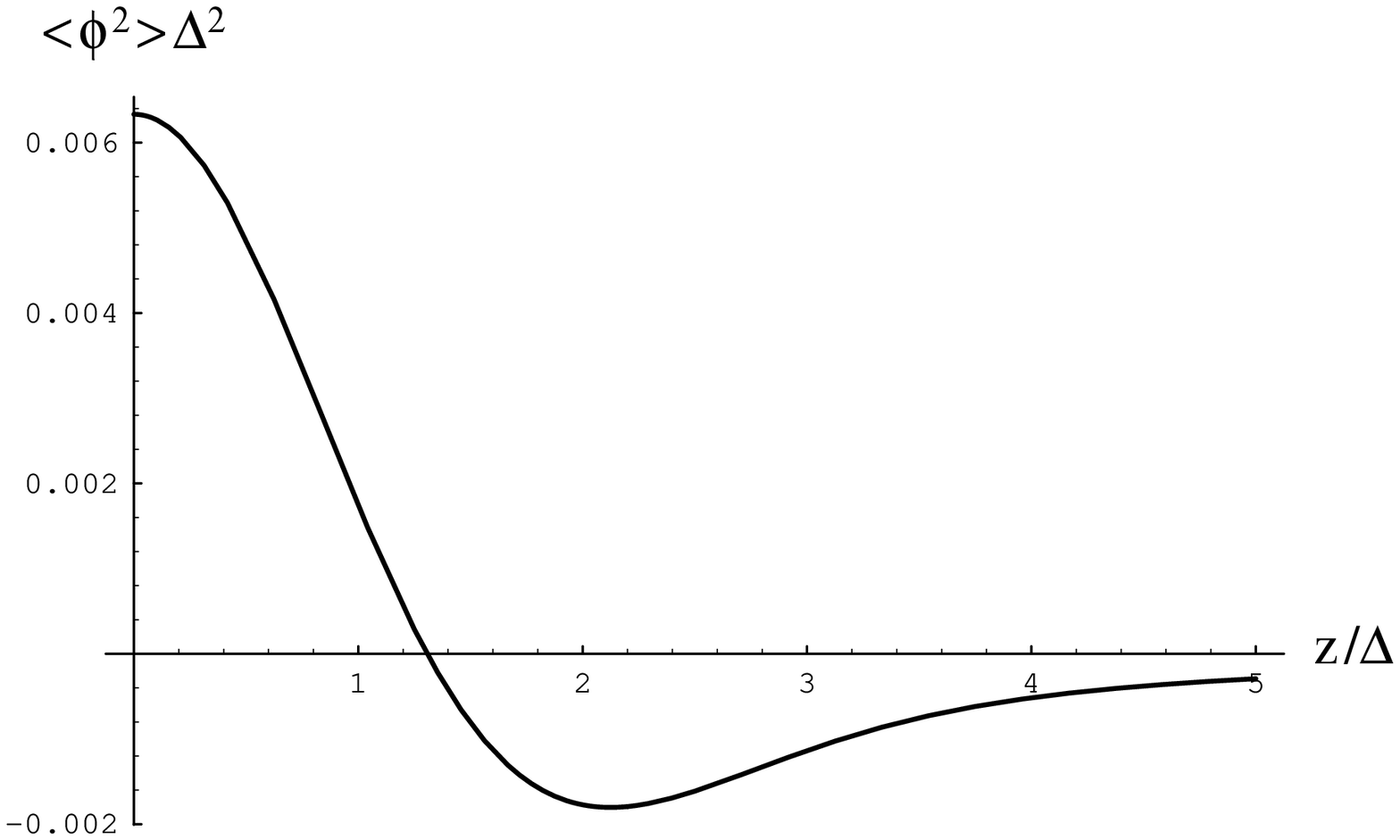}
\label{Figure 1}
\end{center}
\begin{caption}[]

The mean value of $\varphi^2$ near a mirror undergoing 
Gaussian position fluctuations is shown.  Here $\Delta$ is the characteristic
width of the probability distribution.  For $z/\Delta$ large, $\langle 
\varphi^2 \rangle$ is approximately given by Eq.~(\ref{eq:phisq}), but
it is finite as $z/\Delta\rightarrow 0$.
\end{caption}
\end{figure}

\section{The Energy Density near a Single Plate}
\label{sec:rho1}
\subsection{Gaussian Fluctuations}
\label{sec:gaussian}

We now wish to apply the procedure used in the previous section to find 
the energy density in the presence of a single plate, whose position is 
undergoing fluctuations with a Gaussian probability distribution. This
energy density may be expressed as
\begin{equation}
\langle  \rho \rangle = \frac{1}{2} \, 
\lim_{{t' \to t}\atop {{\bf x'} \to {\bf x}}} \Bigl( \partial_t \,  \partial_t'
+ {\bf \nabla}_{\bf x} \cdot {\bf \nabla}_{\bf x'} \Bigr) 
\langle  G_R(x,x') \rangle \,.  \label{eq:rho1}
\end{equation}
A repetition of the procedure used for $\langle  G_R \rangle$ leads to
\begin{equation}
\langle  \rho \rangle = - \frac{1}{2(2\pi)^3} \,
\lim_{{t' \to t}\atop {{\bf x'} \to {\bf x}}}
\, {\rm Re} \int \frac{d^3 k}{\omega} \, (\omega^2 -k_z^2)\,
 e^{i{\bf k}_t \cdot ({\bf x}_t - {\bf x}'_t)}
\, e^{-i \omega (t-t')} \, e^{i k_z (z+z')} \, \langle e^{-2i k_z q} \rangle 
\, .                                                 \label{eq:rho2}
\end{equation} 
If we now employ the relation for Gaussian fluctuations, Eq.~(\ref{eq:exp}),
and perform the integrations as before, we find
\begin{eqnarray}
 \langle  \rho \rangle                                      
&=& \frac{2 i}{(2\pi)^3\,  \Delta} \, \sqrt{\frac{\pi}{2}}\, \lim_{t' \to t} \,
\frac{\partial^2}{\partial u \,\partial v} \, \biggl\{ \frac{1}{u+v}
\, \biggl[e^{-u^2/(8\Delta^2)} \,  \Phi\left(i \frac{u}{2\sqrt{2} \Delta}\right)
   \nonumber \\
 &+&  e^{-v^2/(8\Delta^2)} \,  
  \Phi\left(i \frac{v}{2\sqrt{2} \Delta}\right)
          \biggr]  \biggr\}  \, ,
\end{eqnarray}
where $u = t-t' -2z$, and $v = t-t' +2z$. Explicit evaluation of the last
expression leads to
\begin{equation}
\langle  \rho \rangle = \frac{1}{192 \,\pi^2 \, \Delta^7}\, \left[
\sqrt{2 \pi}\,  z\,(z^2 -3 \Delta^2)
e^{-z^2/(2\Delta^2)} \, i\, \Phi\left(i \frac{z}{\sqrt{2} \Delta}\right)
+ 2 \Delta (z^2 -2 \Delta^2) \right] \,. \label{eq:rho3}
\end{equation}
Far from the mirror, the energy density is that calculated without 
fluctuations:
\begin{equation}
\langle \rho \rangle \sim - \frac{1}{16\, \pi^2\, z^4}  +\cdots \,, \quad  
z \gg \Delta \, ,
\end{equation}
and near the mirror it is finite
\begin{equation}
\langle \rho  \rangle \rightarrow -\frac{1}{48\, \pi^2\, \Delta^4} \quad  
z \rightarrow 0 \, .
\end{equation}
The energy density, $\langle \rho  \rangle$ is given as a function of $z$
in Fig. 2.

The remaining components of the expectation value of the stress tensor,
$\langle T_{\mu\nu}  \rangle$, may be readily obtained. This must be a Lorentz
tensor formed from the metric $\eta_{\mu\nu}$ and $n_\mu n_\nu$, where
$n^\mu = (0,0,0,1)$ is the unit normal vector to the mirror. Hence, 
 \begin{equation}
\langle T_{\mu\nu}  \rangle = F_1(z)\,\eta_{\mu\nu} + F_2(z)\, n_\mu n_\nu \,,
\end{equation}
where $F_1$ and $F_2$ are scalar functions of $z$. We see immediately that
the transverse components are just minus the energy density:
\begin{equation}
\langle T_{xx}  \rangle = \langle T_{yy}  \rangle = - \langle \rho \rangle \,.
                                                               \label{eq:Txx}
\end{equation}
Furthermore, the conservation law, $\partial^\mu \langle T_{\mu\nu}  \rangle
= 0$ implies that
\begin{equation}
\frac{d}{d z} \bigl[F_1(z) - F_2(z) \bigr] =0 \,.
 \end{equation}
We define the renormalized stress tensor so that $\langle T_{\mu\nu}  \rangle 
\rightarrow 0$ as $z \rightarrow \infty$, which implies $F_1(z) = F_2(z)$,
and hence
 \begin{equation}
\langle T_{\mu\nu}  \rangle = \langle \rho  \rangle \;
                     \bigl( \eta_{\mu\nu} + \, n_\mu n_\nu \bigr)\,.
\end{equation}
As a consequence, the pressure normal to the mirror vanishes:
\begin{equation}
\langle T_{zz}  \rangle = 0 \,.  \label{eq:Tzz}
\end{equation}

Finally, we note that one may calculate the integral of the right-hand-side
of Eq.~(\ref{eq:rho3}) explicitly and verify that
\begin{equation}
\int_0^\infty \langle \rho \rangle \, dz = 0 \,.
\end{equation}
This confirms that the boundary fluctuations remove the apparent discrepancy
between the Casimir energies of the minimal and conformal scalar fields.
Note that although $\langle \rho \rangle$ is negative both at large distances
and at the mean position of the mirror, it is positive in a region near $z=0$,
as illustrated in Fig. 2. The positive energy region can be regarded as 
the concrete realization of the positive surface energy density conjectured
by Kennedy, {\it et al} \cite{KCD}.

\begin{figure}
\begin{center}
\leavevmode\epsfysize=8cm\epsffile{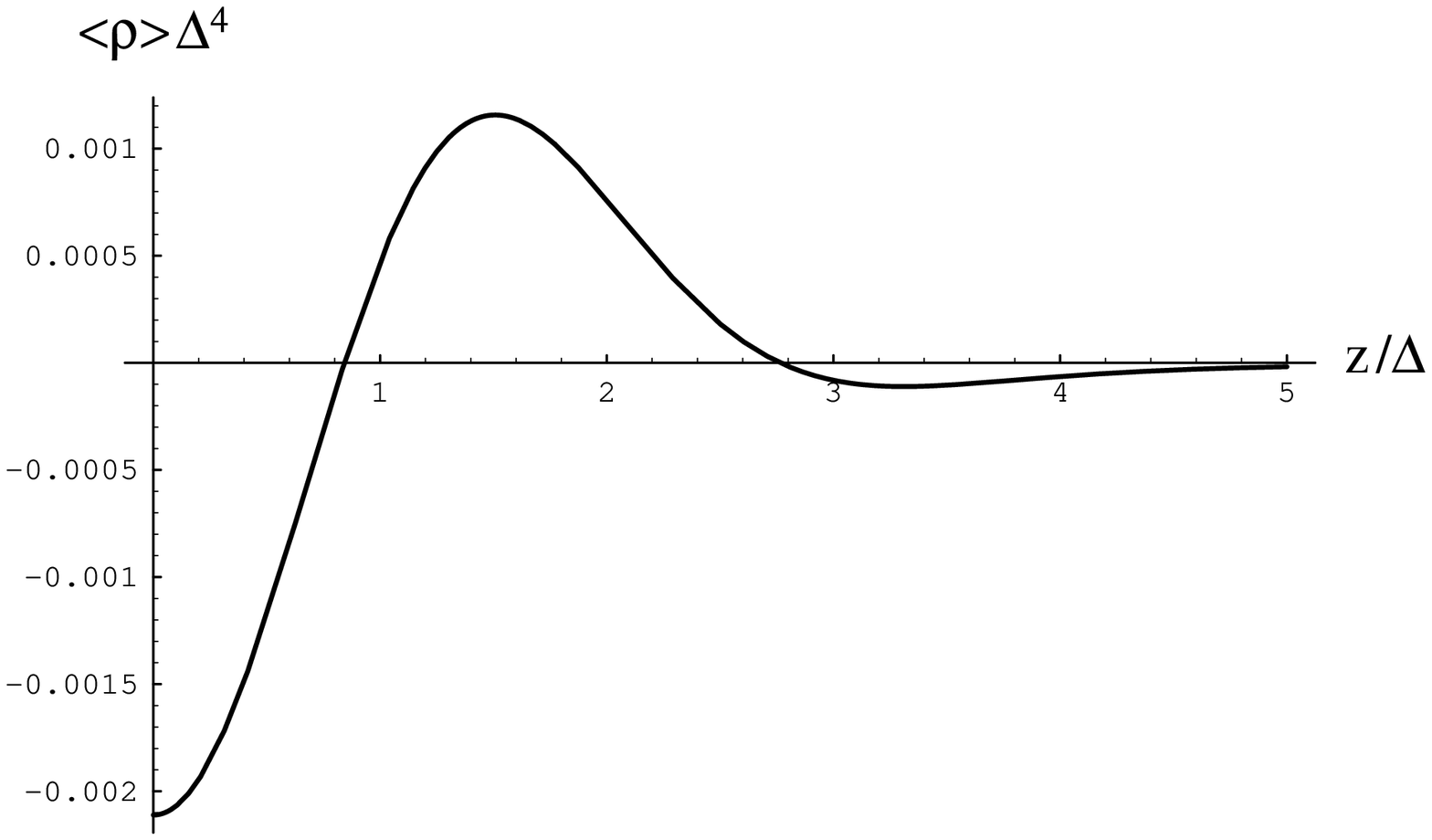}
\label{Figure 2}
\end{center}
\begin{caption}[]

The mean energy density, $\langle \rho \rangle$, near a mirror 
undergoing Gaussian position fluctuations is shown.  The integral of this 
function over all positive $z$ vanishes.
\end{caption}
\end{figure}

\subsection{A General Probability Distribution}

Now we wish to generalize our discussion to an arbitrary probability 
distribution. For later use, we will momentarily assume that the distribution
function $f$ is peaked symmetrically around an arbitrary value of 
$\langle q \rangle$, and write $f=f(s)$, where $s= q - \langle q \rangle$.
The average of a complex exponential function of $q$ then becomes
\begin{equation}
\langle e^{i \alpha q} \rangle =
 e^{i \alpha \langle q \rangle}\, \hat f(\alpha) \, ,
\end{equation}
where $\hat f$ denotes the Fourier transform of $f$:
\begin{equation}
{\hat f}(\alpha) = \int_{-\infty}^\infty  e^{i \alpha s}\, f(s)\, ds \,.
                                                     \label{eq:fourier}
\end{equation}
We may now use Eq.~(\ref{eq:rho2}) to express the averaged energy density for 
an arbitrary, symmetric probability distribution as 
\begin{equation}
\langle \rho  \rangle =  -\frac{1}{2 \pi^2}\, \lim_{t' \to t} \,
\frac{\partial^2}{\partial u \,\partial v} \, Re \,\biggl\{ \frac{i}{u+v} \;
\int_0^\infty d k_z\, \Bigl[ e^{-i k_z u} \, {\hat f}(-2 k_z)
+ e^{-i k_z v} \, {\hat f}(2 k_z) \Bigr] \biggr\} \,,
\end{equation}
where, as before, $u = t-t' -2z$, and $v = t-t' +2z$. We next use 
Eq.~(\ref{eq:fourier}) to re-express $\langle \rho  \rangle$ in terms of $f$,
and employ the relation 
\begin{equation}
\int_0^\infty dx \, e^{i a x} = \frac{i}{a} + \pi\, \delta(a) \label{eq:intexp}
\end{equation}
to find
\begin{equation}
\langle \rho  \rangle = - \frac{1}{\pi^2}\, \lim_{t' \to t} \,
\frac{\partial^2}{\partial u \,\partial v} \, \left[ 
               \frac{F(u) + F(v)}{u+v} \right]  \, , \label{eq:rho4}
\end{equation}
where
\begin{equation}
F(u) =  \int_{-\infty}^\infty d s \, \frac{f(s)}{2s + u} \,,
\end{equation}
and the last integral is understood to be a principal value. 

Thus given the probability distribution $f(s)$, we need only calculate
$F$ (the Hilbert transform of $f$), and then evaluate the derivatives and limit
in Eq.~(\ref{eq:rho4}). It is of interest to apply this formalism to the
case of a compactly-supported distribution. A simple example is
\begin{equation}
f(s) = \frac{315}{256\, s_0^9}\, (s -s_0)^4\, (s +s_0)^4 \, ,
                        \quad -s_0 \leq s \leq s_0 \,,  \label{eq:compact}
\end{equation}
and $f(s) = 0$ for $|s| > s_0$. This function is chosen so that $f$ and its
first three derivatives are continuous at $s = \pm s_0$. 
Equation~(\ref{eq:rho4}) now leads to
\begin{eqnarray}
\langle \rho  \rangle &=& \frac{1}{512\,\pi^2\, s_0^9}\; \left\{
(2205 z^5 -3150 s_0^2 z^3 +945 s_0^4 z) \Bigl[ \ln(z+s_0) - \ln|z-s_0| \Bigr]
  \right. \nonumber \\
  && \left. -4410 s_0 z^4 +4830 s_0^3 z^2 -672 s_0^5 \right\}\,.\label{eq:rho5}
\end{eqnarray}
This function is plotted in Fig. 3. Again, we have that at large distances
from the mirror
\begin{equation}
\langle \rho \rangle \sim - \frac{1}{16\, \pi^2\, z^4}  +\cdots \,, \quad  
z \gg s_0 \, .
\end{equation}
Note that $\langle \rho  \rangle$ has a cusp at $z =s_0$. If we had chosen
a distribution function for which any of the first three derivatives are
discontinuous at this point, $\langle \rho  \rangle$ would become singular
there. Similarly, we could smooth out the cusp by matching more derivatives
at this point. 
We may verify directly that the total energy again vanishes:
\begin{equation}
\int_0^\infty \langle \rho \rangle \, dz = 0 \,.
\end{equation}      
\begin{figure}
\begin{center}
\leavevmode\epsfysize=8cm\epsffile{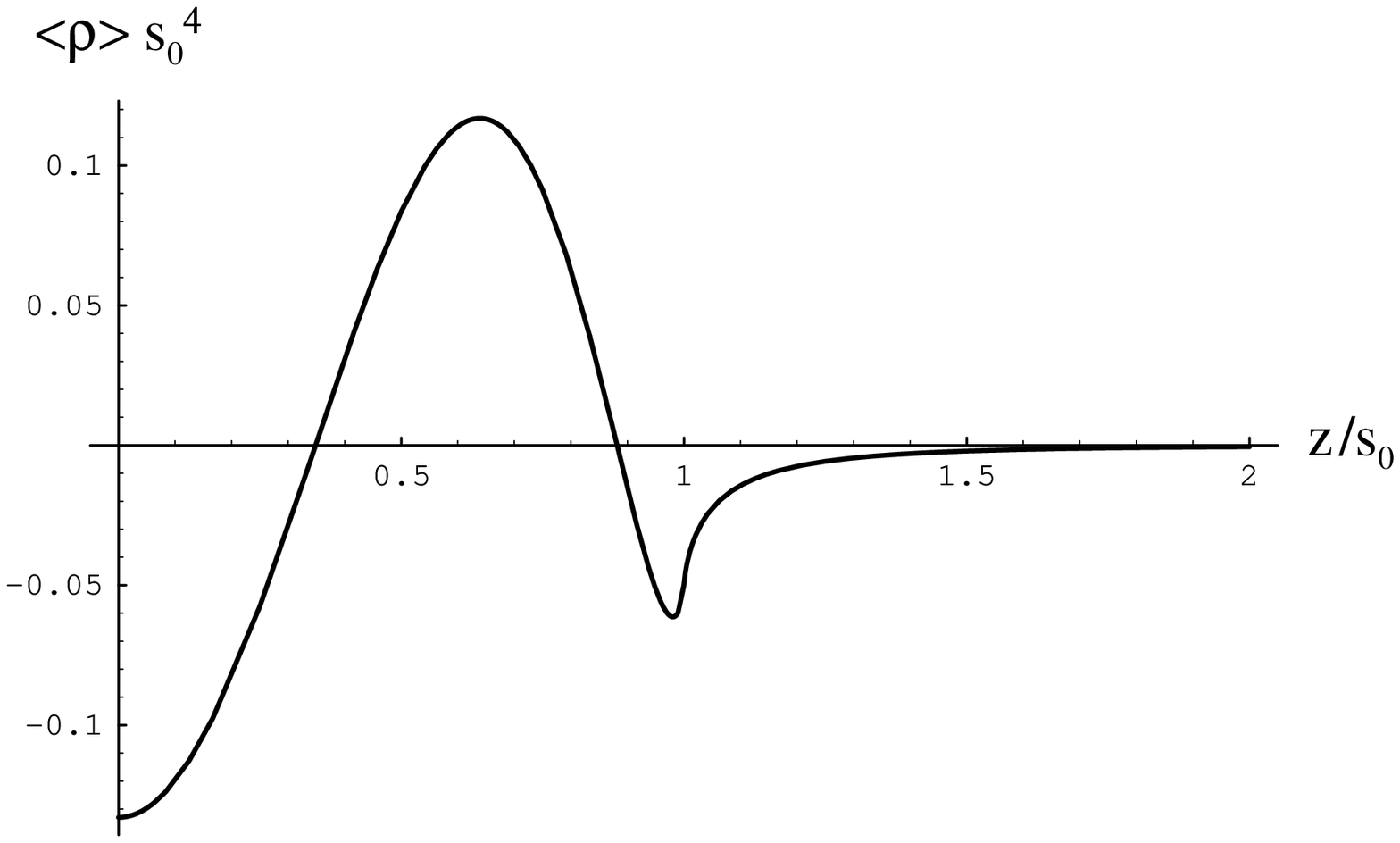}
\label{Figure 3}
\end{center}
\begin{caption}[]

The mean energy density, $\langle \rho \rangle$, is illustrated
for the compact probability distribution, Eq.~(\ref{eq:compact}),
where $s_0$ is the width of the probability distribution.  Again,
the integral of $\langle \rho \rangle$ over positive $z$ vanishes.
\end{caption}
\end{figure}

\section{The Squared Electric Field near a Single Plate}
\label{sec:Esq}

As noted in the Introduction, the squared electric and magnetic fields
diverge in the presence of a perfectly reflecting plate with no position
fluctuations. Specifically, one has
\begin{equation}
\langle {\bf E}^2 \rangle = - \langle {\bf B}^2 \rangle =
\frac{3}{16\, \pi^2\, z^4} \, ,
\end{equation}
so the energy density vanishes:
\begin{equation}
\langle \rho \rangle = \frac{1}{2} \Bigl(\langle {\bf E}^2 \rangle +
\langle {\bf B}^2 \rangle \Bigr) = 0 \,.
\end{equation}
We may now calculate the squared electric field in the presence of a 
fluctuating boundary.  The two point function for the photon field is
\begin{equation}
D^{\mu \nu} (x,x') = \langle 0|  A^\mu (x)\, A^\nu (x') |0 \rangle \,.
\end{equation}
In the presence of the reflecting plate, this may be expressed as
\begin{equation}
{D^{\mu \nu}} (x,x') = {D^{\mu \nu}_{0}} (x-x') +{D^{\mu \nu}_{R}} (x,x')
\end{equation}
where ${D^{\mu \nu}_{0}} (x-x')$ is the two point function in the absence of 
the boundary, and the renormalized two point function, 
${D^{\mu \nu}_{R}} (x,x')$, is the correction introduced by the presence of 
the boundary. In a particular choice of gauge, we have that
\begin{equation}
{D^{\mu \nu}_{0}} (x-x') = {\eta^{\mu\nu}}{G_0}(x-x') \, ,
\end{equation}
and 
\begin{equation}
D_{R}^{\mu \nu} (x,x') = -({\eta^{\mu\nu} + 2{n^\mu}{n^\nu}})
{G_R}(x, x')\,. \label{eq:rgf2}
\end{equation}
Here $G_0$ and $G_R$ are the scalar two point functions given in 
Eqs.~(\ref{eq:gf}) and (\ref{eq:rgf}), respectively, and $n^\mu = (0,0,0,1)$ 
is the unit vector normal to the plate. The renormalized field strength  
two point function can now be obtained by taking the four dimensional curl
in $x$ and in $x'$ of Eq.~(\ref{eq:rgf2}). The electric field part of this
function is 
\begin{equation}
\langle E_i(x)\, E_j(x')  \rangle = \langle F_{0i}(x)\, F_{0j}(x')  \rangle
= \partial_0 \partial_0' \langle A_i(x)\, A_j(x')  \rangle  +
\partial_i \partial_j' \langle A_0(x)\, A_0(x')  \rangle \,.
\end{equation}
In the presence of boundary fluctuations, the mean squared electric field
is now given by
\begin{equation}
\langle  {\bf E}^2 \rangle =  
\lim_{{t' \to t}\atop {{\bf x'} \to {\bf x}}} \Bigl( \partial_t \,  \partial_t'
- {\bf \nabla}_{\bf x} \cdot {\bf \nabla}_{\bf x'} \Bigr) 
\langle  G_R(x,x') \rangle = \frac{1}{3} \langle T^\mu_\mu \rangle \,,  
                                                            \label{eq:Esq}
\end{equation}
where $T^\mu_\mu$ is the trace of the minimal scalar field stress tensor.
From Eqs.~(\ref{eq:Txx}) and (\ref{eq:Tzz}), we find that
\begin{equation}
\langle  {\bf E}^2 \rangle = -3 \langle  \rho \rangle \,. \label{eq:Esq2}
\end{equation}
Thus our explicit results for the scalar field energy density, 
$\langle  \rho \rangle$, Eqs.~(\ref{eq:rho3})
and (\ref{eq:rho5}), also give us $\langle  {\bf E}^2 \rangle$ for the
Gaussian and compact probability distributions, respectively. 

It is of interest to note that the Casimir-Polder  potential between a 
polarizable
particle with a frequency-independent polarizability $\alpha$ and a conducting
boundary is given by
\begin{equation}
V(z) = - \frac{1}{2} \, \alpha \,\langle  {\bf E}^2 \rangle \,.
\end{equation}
Thus Figs. 2 and 3 are also plots of $2 V(z)/(3 \alpha)$ for the 
Gaussian and compact probability distributions, respectively. In both cases,
there is a minimum in $V(z)$ at a finite distance from the mean position
of the boundary, at which the particle could apparently become trapped.
This should probably not be taken too seriously. In the case of Gaussian
fluctuations there is a nonzero probability to find the mirror to the {\it right}
of the minimum of $V(z)$. In the case of the compact probability distribution,
Eq.~(\ref{eq:compact}), the mirror may be found at any location to the left
of the minimum of $V(z)$.

\section{The Energy Density between Two Plates}
\label{sec:rho2}

Here we will address the problem of finding the mean energy density for a
minimally coupled, massless scalar field between a pair of parallel plates.
As before, the field is assumed to vanish on the plates, but their positions
are allowed to fluctuate. First suppose that the plates are fixed at $z=\eta$
and $z=\sigma$, respectively. The two point function may be constructed as an
image sum:
\begin{eqnarray}
G(x,x') &=& - \frac{1}{4 \pi^2}\, \left[ 
\sum_{m=-\infty}^\infty \frac{1}{\Delta \tau^2 -[z-z' +2m(\sigma-\eta)]^2} 
 \right. \nonumber \\  &-& \left.
\sum_{m=-\infty}^\infty \frac{1}{\Delta \tau^2 -[z+z' +2m\sigma-2(m+1)\eta]^2} 
 \right] \, ,
\end{eqnarray}
where $\Delta \tau^2 = \Delta t^2 - \Delta {\bf x}_t^2 =  \Delta t^2 -
 \Delta x^2 -  \Delta y^2$. 
The renormalized two point function is obtained by subtracting $G_0(x,x')$,
which amounts to omitting the $m=0$ term in the first summation. The result
may be written as 
\begin{eqnarray}
  &G_R(x,x')&  = \frac{1}{2(2\pi)^3} \, {\rm Re} \int 
\frac{d^3 k}{\omega} \, e^{i{\bf k}_t \cdot \Delta{\bf x}_t}
\, e^{-i \omega \Delta t}  \nonumber \\
&\times& \left[{\sum_{m=-\infty}^\infty}' e^{i k_z [z-z' +2m(\sigma-\eta)]} \,  
- \sum_{m=-\infty}^\infty e^{i k_z [z+z' +2m\sigma-2(m+1)\eta]} \right] \,,
\end{eqnarray}
where the prime on a summation denotes that the $m=0$ term is omitted.

We will assume that the positions of both plates are described by the same 
probability distribution, $f$. We then have that
\begin{eqnarray}
\langle e^{i k_z [z-z' +2m(\sigma-\eta)]} \rangle &=& \int d\eta \, f(\eta) \,
\int d\sigma \, f(\sigma) \; e^{i k_z [z-z' +2m(\sigma-\eta)]} 
                  \nonumber \\     
&=& e^{i k_z [z-z' +2m(\langle \sigma \rangle -\langle \eta\rangle)]} \,
{\hat f}^2(2mk_z) \, ,
\end{eqnarray}
and 
\begin{equation}
\langle e^{i k_z [z+z' +2m\sigma -2(m+1)\eta]} \rangle =
e^{i k_z [z+z' +2m \langle \sigma \rangle -2(m+1)\langle \eta\rangle]} \;
{\hat f}(2mk_z) \, {\hat f}\Bigl(2(m+1)k_z\Bigr) \,.
\end{equation}
The mean energy density is given by Eq.~(\ref{eq:rho1}). If we combine this
expression with the above results, we find
\begin{eqnarray}
&&\langle \rho \rangle = - \frac{1}{8 \pi^2}\, \lim_{\Delta t \rightarrow 0}
 Re \left[ \frac{\partial^2}{\partial t^2 } \Bigl(\frac{-i}{\Delta t}\Bigr)
\int_{-\infty}^\infty dk_z\,  e^{i |k_z| \Delta t} {\sum_{m=-\infty}^\infty}'
e^{2i |k_z| ma} \, {\hat f}^2(2mk_z) + \right.  \nonumber \\
&&  \left. \!\!\!\!\!\!\!\!\!\!\!\! \!\!\!\!\!\!\! 
\left(\frac{1}{4} \frac{\partial^2}{\partial z^2 } - 
\frac{\partial^2}{\partial t^2 } \right) \Bigl(\frac{-i}{\Delta t}\Bigr)
\int_{-\infty}^\infty dk_z\,  e^{i |k_z| \Delta t} {\sum_{m=-\infty}^\infty}
e^{2i |k_z| (ma + z)} \, {\hat f}(2mk_z) \, 
{\hat f}\Bigl(2(m+1)k_z\Bigr) \right]\,,
\end{eqnarray}
where we have set $\langle \sigma \rangle = 0$ and $a =  \langle \eta\rangle$,
so $a$ is the mean separation between the plates. 

We now employ Eq.~(\ref{eq:intexp}) and the identity \cite{jolley}
\begin{equation}
{\sum_{m=-\infty}^\infty} \frac{1}{m - a} = -\pi \cot \pi a \,.
\end{equation}
After the derivatives have been evaluated, our final result may be expressed as
\begin{equation}
\langle \rho \rangle = \langle \rho \rangle_1 + \langle \rho \rangle_2 \,,
\end{equation}
where
\begin{equation}
\langle \rho \rangle_1 = -\frac{\pi^2}{1440}\,  \int_{-\infty}^\infty ds
\int_{-\infty}^\infty dr \, \frac{f(s) f(r)}{(s+r+a)^4} \, , \label{eq:r1}
\end{equation}
and 
\begin{equation}
\langle \rho \rangle_2 = -\frac{\pi^2}{48}\, \int_{-\infty}^\infty ds
\int_{-\infty}^\infty dr \, f(s) f(r) \,
\frac{2 \sin^2\left[\frac{\pi(z+r)}{s+r+a}\right] -3}
     {(s+r+a)^4\, \sin^4\left[\frac{\pi(z+r)}{s+r+a}\right] }  \,. \label{eq:r2}
\end{equation}

Let us first discuss the limit in which the position of both plates is
precisely defined. In this case, we take $f(s) = \delta(s)$ and obtain
\begin{equation}
\langle \rho \rangle_1 = -\frac{\pi^2}{1440\, a^4} \, ,
\end{equation}
and 
\begin{equation}
\langle \rho \rangle_2 = \frac{\pi^2}{48}\, 
\frac{2 \sin^2\left(\frac{\pi z}{a}\right) -3}
     {a^4\, \sin^4 \left(\frac{\pi z}{a}\right) }  \,.
\end{equation}
This is just the usual result \cite{fulling}; $\langle \rho \rangle_1$ is now
the energy density for a conformal scalar field, and $\langle \rho \rangle_2$
diverges on the boundaries.

Now suppose that we take $f(s)$ to be a function with a finite width, and whose
first three derivatives are finite. The integrals in Eqs.~(\ref{eq:r1}) and 
(\ref{eq:r2}) contain poles in the ranges of integration, but the integrals
are well-defined as principal value or generalized principal value
integrals \cite{Davies}. That is, we use identities of the form
\begin{equation} 
\int_{-\infty}^\infty ds \, \frac{f(s)}{(s-a)^4} =
- \frac{1}{3} \int_{-\infty}^\infty ds \, \frac{f'(s)}{(s-a)^3} =
\frac{1}{6} \int_{-\infty}^\infty ds \, \frac{f''(s)}{(s-a)^2} =
- \frac{1}{6} \int_{-\infty}^\infty ds \, \frac{f'''(s)}{s-a} \, .
\end{equation}
Thus we see that both $\langle \rho \rangle_1$ and $\langle \rho \rangle_2$
will be finite everywhere. 

We may now integrate the finite energy density on $z$ to obtain the
mean energy per unit area:
\begin{equation}
E = \int_0^a \langle \rho \rangle \, dz = E_1 + E_2 \, ,
\end{equation}
where 
\begin{equation}
E_1 = a \langle \rho \rangle_1 \, ,
\end{equation}
and 
\begin{equation}
E_2 =  \int_{-\infty}^\infty ds
\int_{-\infty}^\infty dr \, f(s) f(r) \, H(s,r) \, ,
\end{equation}
and $H(s,r)$ is defined by
\begin{equation}
H(s,r) = -\frac{\pi}{48}\, \frac{ \cos\left[\frac{\pi r}{s+r+a}\right] \,
                                      \sin^3\left[\frac{\pi(z+r)}{s+r+a}\right]
- \sin^3\left[\frac{\pi r}{s+r+a}\right] \,
                                    \cos\left[\frac{\pi(z+r)}{s+r+a}\right] }
{(s+r+a)^3 \,\sin^3\left[\frac{\pi r}{s+r+a}\right]\, 
             \sin^3\left[\frac{\pi(z+r)}{s+r+a}\right] } \, .
\end{equation}
In order to discuss the case of plates which are highly localized in position, 
we need to Taylor expand  $H(s,r)$ around $s=r=0$:
\begin{equation}
H(s,r) \approx - \frac{1}{48 \pi^2}\, 
\left( \frac{1}{s^3} + \frac{1}{r^3} \right) + \frac{\pi^2 (s + r)}{720 a^4}
 + \cdots \,.
\end{equation}
Although the leading term in this expansion is singular at $s=0$ or $r=0$,
its contribution to $E_2$ vanishes because of the symmetry of the probability
distribution, $f(s) = f(-s)$. All subsequent terms in the expansion of $H(s,r)$
vanish at $s=r=0$. Thus we find that if we first form the total energy per
unit area of the plates, and then take the limit in which $f(s) \rightarrow 
\delta(s)$, the result is the same as for the conformal scalar field:
\begin{equation}
E_1 \rightarrow -\frac{\pi^2}{1440\, a^3} \,, \qquad
E_2 \rightarrow 0 \,.
\end{equation}
In both cases, we now find the same, negative Casimir energy. (Note that,
in general, the sign of a Casimir energy is very difficult to predict in
advance of an explicit calculation, and can depend upon both boundary
conditions and the dimensionality of spacetime \cite{CPSS}.) 

\section{Summary and Conclusions}
\label{sec:sum}

In the previous sections, we have seen that position fluctuations of a 
reflecting boundary are capable of removing divergences in the renormalized
values of local observables, such as $\langle \varphi^2 \rangle$ and
$\langle T_{\mu\nu} \rangle$. In the case of the massless, minimally coupled
scalar field, such fluctuations also remove the discrepancy between 
the spatial integral of the renormalized energy density, 
$\int \langle T_{tt}\, \rangle d^3x$, and the renormalized expectation value
of the Hamiltonian, $\langle H \rangle$. Position fluctuations are necessary
if one is to treat the mirror as a quantum mechanical object.

Of course, for real mirrors the mass is likely to be so large that the
position uncertainty $\Delta$ is very small. In this case, the cutoff in
reflectivity due to dispersion will normally be the dominant effect. Dielectric
materials become transparent to electromagnetic radiation at wavelengths
shorter than about the plasma wavelength, $\lambda_p$. So long as the position
uncertainty is small compared to this length, $\Delta \ll \lambda_p$, dispersive
effects are dominant, and the position fluctuations may be ignored. However, 
if one could arrange to prepare a mirror in a quantum state in which 
$\Delta > \lambda_p$, the position fluctuation effects discussed in this
paper would become dominant.

In the area of gravitational physics, the situation is rather different.
Here it is also possible to have horizons which act as boundaries for
quantized fields, and for the renormalized expectation value of the stress
tensor to diverge on the horizon. As discussed in the Introduction,
examples include the event horizon of Schwarzschild
spacetime in the Boulware vacuum, the Cauchy horizon in Misner space,
and possibly the chronology horizon in a spacetime containing closed
timelike curves. Now there is no natural cutoff at high frequencies, and
in fact higher frequency modes tend to couple more strongly to gravity
by virtue of their larger energy. It is of course possible that a more
complete quantum theory of gravity will introduce an effective cutoff
at the Planck scale. At the present, any discussion of Planck scale physics
must be regarded as highly speculative. Nonetheless, position fluctuations
of the horizon would seem to provide a possible way to avoid divergent 
stress tensors. It is plausible that the location of a spacetime horizon
undergoes position fluctuations due either to the quantum nature of gravity
(``active fluctuations'') \cite{FS97}, or to fluctuations  of the stress tensor
of quantum matter fields (``passive fluctuations'') \cite{sorkin}. 
This is a topic requiring further study.
 
\vskip .8cm
{\bf Acknowledgements:} We would like to thank Bruce Jensen and Michael
Pfenning for comments on the manuscript. This work was
supported in part by the National Science Foundation (Grant No. PHY-9507351)
and by Conselho Nacional de
Desevolvimento Cientifico e Tecnol{\'o}gico do Brasil (CNPq).


\begin{thebibliography}{--}
\bibitem{fulling} See, for example, S. A. Fulling, {\it Aspects of 
                          Quantum Field Theory in Curved Spacetime}, 
                          (Cambridge University Press, Cambridge, England, 
                          1989),  p. 105-106. 

\bibitem{DC} D. Deutsch and P. Candelas, Phys. Rev. D {\bf 20}, 
                      3063 (1979); P. Candelas, Ann. Phys. (N.Y.)
{\bf 143}, 241 (1982).

\bibitem{KCD} G. Kennedy, R. Critchley, and S. Dowker, Ann. Phys. (N.Y.)
{\bf 125}, 346 (1980).

\bibitem{SS1}  N.F. Svaiter and B.F. Svaiter, J. Math. Phys. {\bf 32}, 175
(1991).

\bibitem{SS2}  N.F. Svaiter and B.F. Svaiter, J. Phys. A {\bf 25}, 979
(1992).

\bibitem{Unruh} W.G. Unruh, Phys. Rev. D {\bf 14}, 870 (1976);
{\bf 15}, 365 (1977).

\bibitem{visser} M. Visser, Phys. Rev. D {\bf 54}, 5116 (1996), gr-qc/9604008.

\bibitem{HK} W.A. Hiscock and D.A. Konkowski, Phys. Rev. D {\bf 26}, 
                1225 (1982).
\bibitem{F95} L.H. Ford, Phys. Rev. D {\bf 51}, 1692 (1995).

\bibitem{HCP} S.W. Hawking, Phys. Rev. D {\bf 46}, 603 (1992).

\bibitem{jolley} L.B.W. Jolley, {\it Summation of Series}, (Dover, New York,
1961), p 100.

\bibitem{Davies} K.T.R. Davies and R.W. Davies,
Can. J. Phys. {\bf 67}, 759 (1989); K.T.R. Davies, R.W. Davies,
and G. D. White, J. Math. Phys. {\bf 31}, 1356 (1990).

\bibitem{CPSS} See, for example, F. Caruso, N. Pinto-Neto, B.F. Svaiter,
 and N.F. Svaiter, Phys. Rev. D {\bf 43}, 1300 (1991).

\bibitem{FS97} L.H. Ford and N.F. Svaiter, Phys. Rev. D {\bf 56}, 2226 (1997),
gr-qc//9704050.

\bibitem{sorkin} R.D. Sorkin, {\it Two Topics concerning Black Holes: 
Extremality of the
Energy, Fractality of the Horizon}, gr-qc/9508002; {\it How Wrinkled is the
Surface of a Black Hole?}, gr-qc/9701056; A. Casher, F. Englert, N. Itzhaki, and R. Parentani,
{\it Black Hole Horizon Fluctuations}, Nucl. Phys. {\bf B484}, 419 (1997),
 hep-th/9606106. Note, however,
criticisms of these papers in Ref. \cite{FS97}.



\end{thebibliography}
\end{document}